\documentclass[12pt]{iopart}
\usepackage{iopams}
\usepackage{epsfig}
\usepackage{textcomp}
\bibstyle{plain}
\begin{document}

\title{Evolution of cooperation on scale-free networks subject to error and attack}

\author{Matja{\v z} Perc}
\address{Department of Physics, Faculty of Natural Sciences and Mathematics, University of \\ Maribor, Koro{\v s}ka cesta 160, SI-2000 Maribor, Slovenia}
\ead{matjaz.perc@uni-mb.si}

\begin{abstract}
We study the evolution of cooperation in the prisoner's dilemma and the snowdrift game on scale-free networks that are subjected to intentional and random removal of vertices. We show that, irrespective of the game type, cooperation on scale-free networks is extremely robust against random deletion of vertices, but declines fast if vertices with the maximal degree are targeted. In particular, attack tolerance is lowest if the temptation to defect is largest, whereby a small fraction of removed vertices suffices to decimate cooperators. The decline of cooperation can be directly linked to the decrease of heterogeneity of scale-free networks that sets in due to the removal of high degree vertices. We conclude that the evolution of cooperation is characterized by similar attack and error tolerance as was previously reported for information readiness and spread of viruses on scale-free networks.
\end{abstract}
\pacs{02.50.Le, 89.75.Fb, 89.75.Hc}

\maketitle

Social dilemmas emerge if individual success calls for actions that harm collective wellbeing. While cooperation is widely accepted as the rational strategy leading away from the impeding decline, its evolution in groups of selfish individuals is puzzling \cite{axelrod84}. Evolutionary game theory is frequently used as the framework within which answers to the puzzle are sought \cite{hofbauer88}. The pivotal study that launched a spree of activity aimed towards resolving dilemmas of cooperation in evolutionary games is due to Nowak and May \cite{nowaknat92}, who showed that spatial structure may, unlike well mixed populations, maintain cooperative behavior in the prisoner's dilemma game. However, it has also been reported that spatial structure may inhibit cooperation in the snowdrift game \cite{hauertnat04}, and thus the need for finding a unifying mechanism supporting the cooperative strategy was revitalized. Quite remarkably, scale-free networks \cite{barabasisci99} turned out to be the missing link to cooperation by virtually all main social dilemmas \cite{santosprl05, santospnas06}, owning predominantly to the heterogeneity that characterizes their degree distribution. Although several studies have since elaborated on different aspects of cooperation on scale-free networks, as for example dynamical organization \cite{gardenesprl07, puschpre08}, clustering \cite{assenzapre08} and mixing patterns \cite{rongpre08}, as well as memory \cite{wangpre08}, robustness \cite{poncelanjp07} and payoff normalization \cite{santosjeb06, masudaprsb07, szolnokipa08}, open questions still remain. Notably, the promotive impact of heterogeneous states on the evolution of cooperation has been reported also in other contexts \cite{szolnokiepl07, percpre08, santosnat08}, and coevolutionary rules have been introduced that may generate appropriate heterogeneities spontaneously \cite{pachecoprl06, lipre07, szolnokinjp08, szolnokiepl08, fupre08}.

Importantly though, complex networks received outstanding attention already substantially before above findings were published \cite{albertrmp02, newmansiam03, boccalettipr06}, and indeed, it can be stated that evolutionary games, apart from some notable exceptions \cite{abramsonpre01, ebelpre02, szolnokipre04}, were actually late in getting on board. For a comprehensive review see \cite{szabopr07}. A particular aspect of studies concerning scale-free and related complex networks, that is of great relevance for the present work, is their tolerance to attack and error \cite{albertnat00}, upon which relies fast availability of information within the world-wide-web \cite{pastorprl01}, uninterrupted supply with electricity \cite{albertpre04}, fast spread of epidemics and viral infections \cite{pastorpre02, zanettepa02, barthelemyprl04, colizzaplos07}, robust and near flawless reproduction of organisms \cite{hartwellnat99}, and surely many other aspects of everyday life. Evidently, some of these features are very desirable, while others we would be better off without. Remarkable is that the deletion of high degree vertices, constituting an intentional attack on the network, severely impairs all these processes, whereas random vertex removals, constituting what can be interpreted as errors in communication pathways across the network, leave them practically unaffected. Due to the obvious importance of the subject, several studies elaborated on the resilience of complex networks also analytically via the usage of percolation theory \cite{cohenprl00, callawayprl00, cohenprl01, pietschpre06}.

Aim of the presented work is to show that the evolution of cooperation on scale-free networks obeys to similar laws as mentioned above, in that the removal of high degree vertices rapidly decimates the density of cooperators, whereas random deletion has a completely negligible impact. More precisely, we report that the impact of attack depends significantly on the temptation to defect, whereas errors go by unnoticed irrespective of game parametrization. We will also show that these results can be directly attributed to the rapid decrease of degree heterogeneity following an intentional attack on the scale-free network, and the lack thereof in case of random errors. The study thus supplements previous works examining the evolution of cooperation on scale-free networks, and more importantly, intimately links the process of strategy evolution in terms of error and attack tolerance to earlier studies on epidemic spread and information propagation.

In what follows, both the prisoner's dilemma game as well as the snowdrift game will be used as representative examples of social dilemmas, whereby we adopt the same parametrization as used recently in \cite{santosprl05}. Accordingly, the prisoner's dilemma game is characterized by the temptation to defect $T=b$, reward for mutual cooperation $R = 1$, and punishment $P$ as well as the suckers payoff $S$ equaling $0$, whereby $1 < b \leq 2$ ensures a proper payoff ranking \cite{nowaknat92}. The snowdrift game, on the other hand, has $T=\beta$, $R=\beta-$\textonehalf, $S=\beta-1$ and $P=0$, where the temptation to defect can be expressed in terms of the cost-to-benefit ratio $r=1/(2\beta -1)$ with $0 \leq r \leq 1$. In both games two cooperators facing one another acquire $R$, two defectors get $P$, whereas a cooperator receives $S$ if facing a defector who then gains $T$. Initially, each player $x$ on the network is designated either as a cooperator $(C)$ or defector $(D)$ with equal probability. Irrespective of the game, evolution of the two strategies is performed in accordance with the Monte Carlo simulation procedure comprising the following elementary steps. First, a randomly selected player $x$ acquires its payoff $p_x$ by playing the game with all its $k_x$ neighbors. Next, one randomly chosen neighbor of $x$, denoted by $y$, also acquires its payoff $p_y$ by playing the game with all its $k_y$ neighbors. Last, if $p_x > p_y$ player $x$ tries to enforce its strategy $s_x$ on player $y$ in accordance with the probability $W(s_x \rightarrow s_y)=(p_x-p_y)/b k_q$, where $k_q$ is the largest of the two degrees $k_x$ and $k_y$. In accordance with the random sequential update, each player is selected once on average during a full Monte Carlo step.

As the underlying interaction topology we use scale-free networks generated via growth and preferential attachment as proposed by Barab{\'a}si and Albert \cite{barabasisci99}, whereby each vertex corresponds to a particular player $x$. The generation of the network starts with two connected players, and subsequently every new player is attached to two old players already present in the network, whereby the probability $\Pi$ that a new player will be connected to an old player $x$ depends on its degree $k_x$ in accordance with $\Pi=k_x / \sum k_y$. This growth and preferential attachment scheme yields a network with an average degree $\kappa=(1/N) \sum k_x$ of four, and a power-law degree distribution with the slope of the line equaling $-2.9$ on a double logarithmic graph. Notably, analytical estimations predict the slope of the line to equal $-3$ \cite{barabasisci99}.Below we will study the evolution of cooperation in dependence on the fraction of deleted vertices $\Lambda = \eta /N$, where $\eta$ is the number of deletions to be made and $N$ is the original size of the network. As noted, two procedures for the removal of vertices will be considered. During the first, all $\eta$ vertices to be deleted are selected randomly from the network. The second is more deliberate, in that we always look for the vertex with the largest degree within the network and remove it. This is done in a consecutive manner until $\eta$ vertices, each having the largest degree within the network at the time of removal, are deleted. The two procedures correspond to error and attack, respectively, as argued in \cite{albertnat00}. Notably, some players that are not selected for deletion may become disconnected from the network during either of the two procedures since removing a vertex entails deleting all its links as well. Such disconnected players are also removed as soon as they appear, but do not contribute to $\eta$. Note here that completely isolated players cannot have an impact on the evolution of cooperation because at least one partner is needed with whom the game can be played. We will also be interested in quantifying the degree heterogeneity of networks obtained by different $\Lambda$, which can be realized succinctly by calculating the degree variance $\chi = (\varrho-\kappa^2)/ \kappa$, where $\varrho$ is the average of the square of all $k_x$ (note that here $\kappa$ and $\varrho$ are evaluated by taking into account only the $N - \eta$ remaining vertices by a given $\Lambda$).

Presented results were obtained on networks hosting $N=5000$ players by $\Lambda=0$. Equilibrium fractions of cooperators $\rho_C$ were determined within $10^6$ full Monte Carlo steps after sufficiently long transients were discarded. Importantly, since the generation of scale-free networks has inherent random ingredients, which can be additionally amplified by vertex deletions, final results shown below were averaged over $200$ independent runs for each set of parameter values to warrant appropriate accuracy.

\begin{figure}
\begin{center} \includegraphics[width = 11cm]{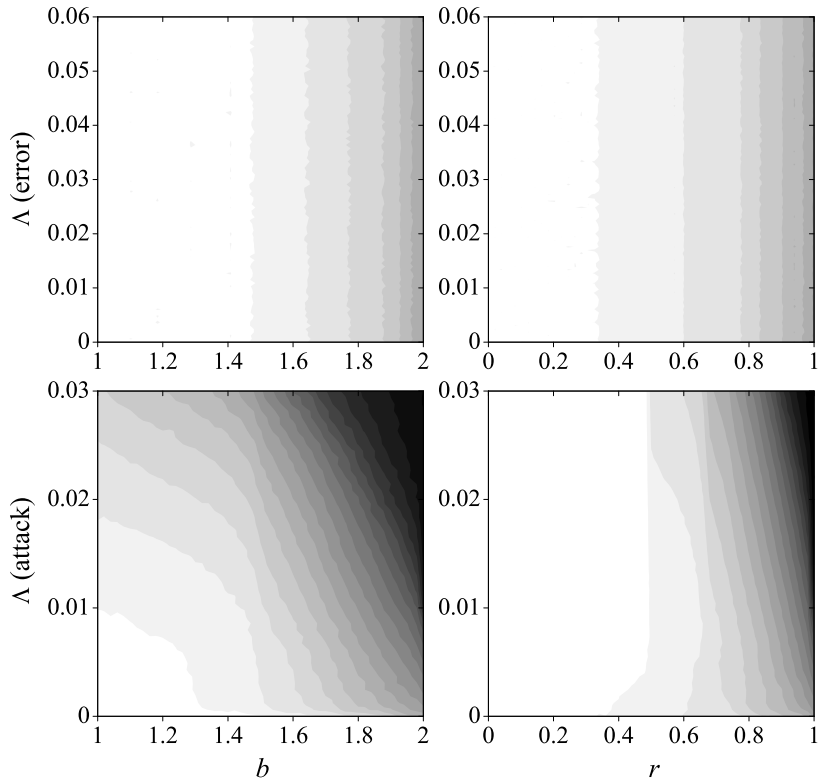}
\caption{\label{fig1} Gray-scale maps of $\rho_C$ in dependence on $b$ (left column; prisoner's dilemma game) and $r$ (right column; snowdrift game) by different $\Lambda$. Top two panels show results for random removal of vertices, whereas bottom two panels depict the impact of attack. Shading in all panels is linear, black depicting $\rho_C=0.1$ and white showing $\rho_C=1.0$ with $20$ steps between the two. In order to strengthen the resilience of $\rho_C$ on error in scale-free networks, the span of $\Lambda$ in the upper two panels is two times larger than in the bottom two panels.}
\end{center}
\end{figure}

Next, we will systematically analyze the impact of error and attack on the evolution of cooperation, whereby results for the prisoner's dilemma and the snowdrift game will be shown and commented in a parallel fashion for the purpose of better comparison options. Upper two panels of Fig.~\ref{fig1} depict the effect of error on the evolution of cooperation within the prisoner's dilemma game (left) and the snowdrift game (right). Irrespective of $\Lambda$, the dependence of $\rho_C$ on different levels of error-affected vertices is negligible by both game types. In fact, it is practically impossible to distinguish the evolution of cooperation on a perfectly functioning scale-free network from the evolution of cooperation taking place on a scale-free network that is prone to error, even if up to $6 \%$ of randomly selected vertices become completely dysfunctional (are deleted along with all their links). This changes rather dramatically if, instead of randomly selected vertices, the targeted vertices become those with the largest degree within the network. Lower two panels of Fig.~\ref{fig1} depict the effect of attack on the evolution of cooperation within the prisoner's dilemma game (left) and the snowdrift game (right). By the prisoner's dilemma game the fraction of cooperators decreases steadily as $\Lambda$ increases, albeit the rate of the decrease differs substantially in dependence on $b$. On the other hand, the detrimental impact on $\rho_C$ by the snowdrift game is virtually absent for $r<0.36$, and even seems to reverse slightly until $r<0.65$, yet towards the $r=1$ limit cooperators are, similarly as by the prisoner's dilemma game, decimated heavily by as low as $3 \%$ of deleted high degree vertices.

\begin{figure}
\begin{center} \includegraphics[width = 10.5cm]{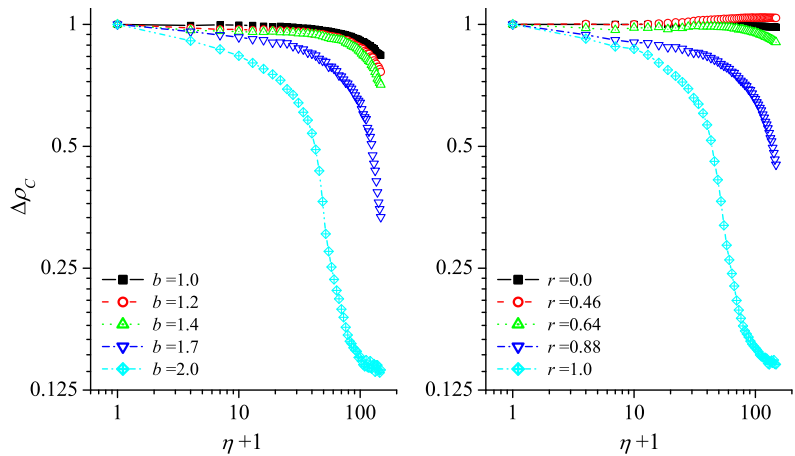}
\caption{\label{fig2} Decrease of $\Delta \rho_C$ (relative to the $\Lambda=0$ case) in dependence on $\eta$ for different values of $b$ (left; prisoner's dilemma game) and $r$ (right; snowdrift game) due to intentional attack. Note that the horizontal axis displays $\eta+1$ so that the reference value can also by displayed on a log scale. Lines are just to guide the eye.}
\end{center}
\end{figure}

In order to present the impact of attack in both games more deliberately, we show in Fig.~\ref{fig2} the relative decrease of cooperator density $\Delta \rho_C$ (with respect to the $\Lambda=0$ case) by representative values of $b$ and $r$. Left panel of Fig.~\ref{fig2} depicts results for the prisoner's dilemma game. As could be inferred already from the bottom left panel of Fig.~\ref{fig1}, the relative decrease in dependence on $\Lambda$ increases fast as $b$ increases, and indeed, by $b=2$ as much as $87 \%$ of cooperators present at $\Lambda=0$ are replaced by defectors at $\Lambda=0.03$ (equivalently $\eta =150$ if $N=5000$). Conversely, by lower values of $b$ the impact of attack may be less devastating, but only if $\Lambda$ remains small. Note that by a substantial deletion of high degree vertices the network approaches a regular graph, and the sustenance of cooperation thus becomes impaired even by low $b$. Results for the snowdrift game, presented in the right panel of Fig.~\ref{fig2}, are largely in agreement with the above interpretation. Differently is, as noticed already by the examination of the bottom right panel of Fig.~\ref{fig1}, that for $r=0.46$ and $r=0.64$ a slight increase in $\Delta \rho_C$ can be inferred by intermediate $\eta$. However, this effect is practically negligible if compared to the subsequent downward trend, and it may be attributed to the differences in the microscopic dynamics of strategy adoptions between the prisoner's dilemma and the snowdrift game, as recently argued in \cite{hauertnat04}. Thus, irrespective of the type of the game, the impact of deliberate vertex deletions becomes increasingly pronounced by high temptations to defect, where as low as $3 \%$ of removed influential players suffice to heavily impair the evolution of cooperation.

\begin{figure}
\begin{center} \includegraphics[width = 9.5cm]{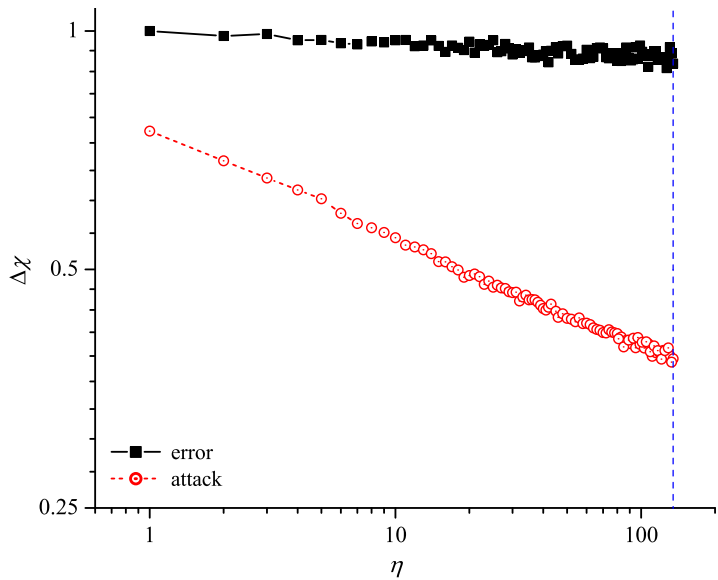}
\caption{\label{fig3} Decrease of $\Delta \chi$ (relative to the $\Lambda=0$ case) in dependence on $\eta$ by error (black squares) and attack (red circles) of scale-free networks. The vertical line (dashed blue) denotes the critical number of removed vertices needed for the disintegration of the considered scale-free network into isolated components \cite{cohenprl01}. Note that both axes have a logarithmic scale. Lines connecting the symbols are just to guide the eye.}
\end{center}
\end{figure}

Above outlined results, presenting the impact of error and attack on the evolution of cooperation, can be explained by studying the resulting degree heterogeneity of networks by different values of $\Lambda$. The outstanding importance of heterogeneity for maintaining cooperative behavior has been described in \cite{santosprl05}, where it has been argued that cooperators may benefit substantially from occupying the hubs of the network, and as doing so spread their strategy effectively. Conversely, defectors are unable to claim lasting benefits from occupying the hubs, simply because they become very weak as soon as all the neighbors of the defecting hub become defectors themselves. At this point influential defectors occupying the hubs become vulnerable and are easily overtaken by cooperators. Similar mechanisms for promotion of cooperation have recently been proposed also for games on regular graphs, where the heterogeneity was introduced via random quenched evolutionary landscapes \cite{percpre08} or differences in the reproduction capability of players \cite{szolnokinjp08}. In these studies it has been shown that as the heterogeneity in the system decreases so does the density of cooperators. Here we quantify heterogeneity by calculating the degree variance $\chi$ of networks in dependence on $\Lambda$ separately for error and attack. Figure~\ref{fig3} shows the relative (with respect to $\Lambda=0$ case) decrease of degree variance $\Delta \chi$ for error (black squares) and attack (red circles). It can be observed that random deletions of vertices have practically no impact on the heterogeneity of resulting networks, whereas attack, on the other hand, decreases the degree heterogeneity in a power-law fashion. In accordance with the previously described impact of heterogeneity on the evolution of cooperation \cite{santosprl05, percpre08, szolnokinjp08, devlinpre09}, we argue that the two depicted dependencies of $\Delta \chi$ on $\Lambda$ in Fig.~\ref{fig3} are responsible for both, the supreme tolerance of cooperation on error, as well as its fragility on attack by high temptations to defect. Importantly, however, the decay of cooperation can also be linked to the critical fraction of removed vertices that is needed for the scale-free network to disintegrate into isolated components. Following the works of Cohen \textit{et al.} \cite{cohenprl00,cohenprl01} based on percolation theory, the critical fraction $\Lambda_c$ can be estimated analytically [see Eqs. (7) and (9) in Ref. 42], in particular $\Lambda_c=0.028 \pm 0.002$. In Fig.~\ref{fig3} the dashed blue vertical line denotes the value of $\eta$ corresponding to $\Lambda_c$, and indeed, it can be inferred that the critical value for network fragmentation agrees well with the maximal decay of cooperation by high temptations to defect (see $b=2.0$ and $r=1.0$ results presented in Fig.~\ref{fig2}; note also the saturation of the decay when approaching $\Lambda_c$). Thus, we conclude that the threshold for network fragmentation can be used as a good estimate for the tolerance of cooperation on scale-free networks subject to attack, and moreover, that the transition to defector dominance can in fact be related to network fragmentation into isolated components.

In sum, we have elaborated on the tolerance of cooperation evolution in the prisoner's dilemma and the snowdrift game on scale-free networks subject to error and attack. We show that, irrespective of the game type, cooperation on scale-free networks is extremely robust against random deletion of vertices, but declines fast upon attack in environments prone to defection. These two facts are attributed to the impact of vertex deletion on the degree heterogeneity of resulting networks, which is unaffected by error but decreases fast in case of attack. Moreover, we were able to link the decay of cooperators with the critical fraction of intentionally removed vertices that is needed for the scale-free network to disintegrate into isolated components, thus coupling our findings with previous considerations of percolation theory upon deliberate attack \cite{cohenprl01}. Presented results support the claim that the evolution of cooperation by high temptations to defect is characterized by the same faint attack tolerance as was previously reported for several other processes on scale-free networks, as for example spread of viral diseases or the effectiveness of information retrieval (for a review see \cite{newmansiam03}). However, by small temptations to defect, the damage on cooperation imposed by attacks is limited due to the fact that heterogeneity is then not of vital importance for the survival of cooperators, as exemplified already in the seminal work concerning games on regular grids \cite{nowaknat92}. From the perspective of experiences we have from real life, the study offers a nice view on why influential individuals, such as kings in the past or presidents and other high ranking sanctioners at present, should be granted the amenities that uphold high levels of their personal security. The fragile tolerance of cooperation, especially in environments that incline individuals towards defective behavior, indicates that such measure are indeed reasonable, and might have an evolutionary origin of which roots can be traced back to the very beginnings of civilization.

\ack
The author acknowledges support from the Slovenian Research Agency (Grant Z1-9629).

\end{document}